\documentclass[pdflatex,sn-apa]{sn-jnl}

\usepackage{amssymb}
\usepackage{tabularray}
\usepackage{tabularx}
\usepackage{makecell}
\usepackage{eurosym}
\usepackage{url}
\usepackage{lipsum}
\usepackage{soul}
\usepackage{hyperref}
\usepackage{array}
\usepackage{multirow}
\usepackage{pdflscape}

\usepackage{graphicx}%
\usepackage{multirow}%
\usepackage{amsmath,amssymb,amsfonts}%
\usepackage{amsthm}%
\usepackage{mathrsfs}%
\usepackage[title]{appendix}%
\usepackage{xcolor}%
\usepackage{textcomp}%
\usepackage{manyfoot}%
\usepackage{booktabs}%
\usepackage{algorithm}%
\usepackage{algorithmicx}%
\usepackage{algpseudocode}%
\usepackage{listings}%
\usepackage{setspace}

\usepackage{soul}

\theoremstyle{thmstyleone}%
\theoremstyle{thmstyletwo}%

\theoremstyle{thmstylethree}%
\begin{document}

\title[Article Title]{Demand-Agnostic Assessment of On-Demand Pooled Transit Services}


\author*[1]{\fnm{Olha} \sur{Shulika}}\email{olha.shulika@uj.edu.pl}

\author[1]{\fnm{Hanna} \sur{Vasiutina}}

\author[1]{\fnm{Michał} \sur{Bujak}}

\author[1]{\fnm{Farnoud} \sur{Ghasemi}}

\author[1]{\fnm{Rafał} \sur{Kucharski}}

\affil*[1]{\orgname{Jagiellonian University}}


\abstract{This study proposes a method to assess the potential of pooled on-demand transit feeder services in urban areas where demand is not yet known. We introduce the 'fraction of demand', reflecting the probability that a resident will use the service. Demand is generated on the distribution of residents' address points at varying demand fraction levels. Through simulations, we match travellers into pooled rides and evaluate the service's potential using three performance indicators (KPIs). We observe how these KPIs change with varying demand fractions and identify the most promising hub for each area. By setting KPI thresholds, we select the optimal combination of area and hub that meets these thresholds at the lowest demand fraction. This approach provides municipalities with a structured tool for pre-deployment evaluation, helping them choose the most suitable areas for launching new services despite the absence of exact demand data. We illustrate its application through a case study in Krakow, ranking 12 pre-selected areas for feeder service deployment.}


\maketitle

\section{Introduction}\label{sec1}

Cities around the world are seeking to reduce car dependency and improve urban liveability through sustainable transport options. A key focus is promoting public transport (PT) and shared mobility services, particularly flexible on-demand systems designed to address first and last-mile challenges. These systems significantly improve the appeal of PT by seamlessly connecting residential areas with transit hubs, encouraging a shift from private vehicles to sustainable modes of transportation  \citep{brown2021can, lesh2013innovative, shen2018integrating}.

First-mile challenges relate to the difficulties residents face in accessing the major PT networks from their homes, particularly in areas with limited or infrequent services. Flexible on-demand mobility services can bridge this gap, making PT more attractive by offering direct and convenient access to transit hubs. Such innovations improve urban transport efficiency and reduce the environmental impact of city travel.

\begin{figure}[!ht]
	\centering
		\includegraphics[width=\linewidth]{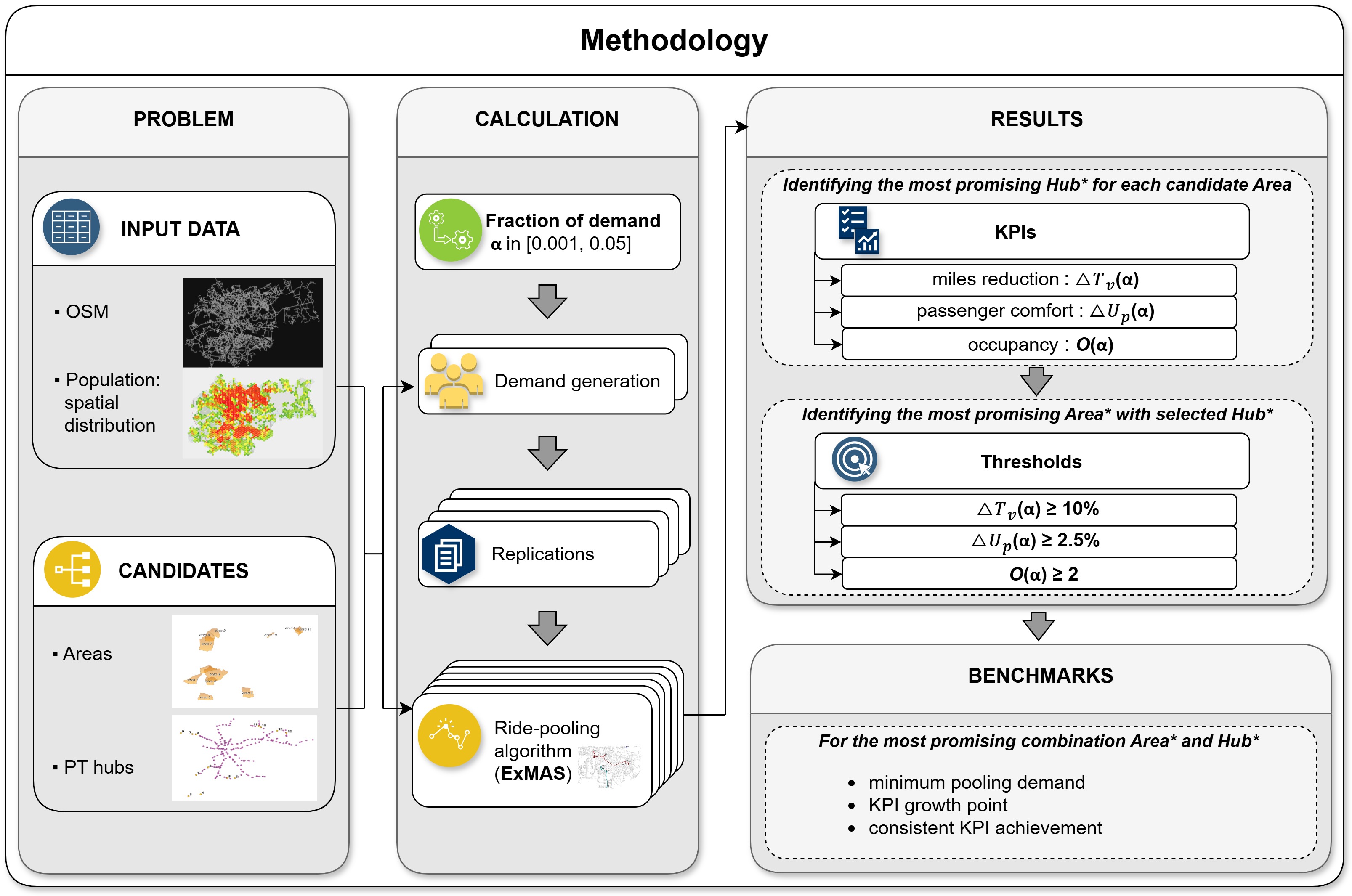}
	\caption {An overview of the applied methodology for selecting a preferred area to implement on-demand pooled transit feeders.}
	\label{FIG:1}    
\end{figure}

Municipalities face several challenges in advancing public transport and shared mobility. A major obstacle is the lack of demand data, which complicates service optimisation \citep{CIVITAS}. Limited funding and competing priorities further restrict investments, especially in resource-constrained regions \citep{vuchic2017urban}. Integrating new on-demand services with existing PT infrastructure requires careful coordination. Gaining public appreciation and the behavioural shift towards shared mobility remain difficult, as many people prefer private vehicles \citep{alonso2020drivers}. Furthermore, regulatory hurdles can block the introduction of flexible, technology-driven services \citep{shaheen2020mobility}. Ensuring equitable access is vital to avoid disparities in mobility provision between populations \citep{manaugh2015integrating}. Addressing these challenges is essential to create accessible and sustainable transport solutions, such as implementing on-demand pooled transit feeders in low-density urban areas where demand data are unavailable. The proposed methodology for addressing this challenge is illustrated in Fig.~\ref{FIG:1}).

\begin{figure}[ht]
    \centering
    \begin{minipage}{0.45\textwidth}
        	\centering
		\includegraphics[width=\linewidth]{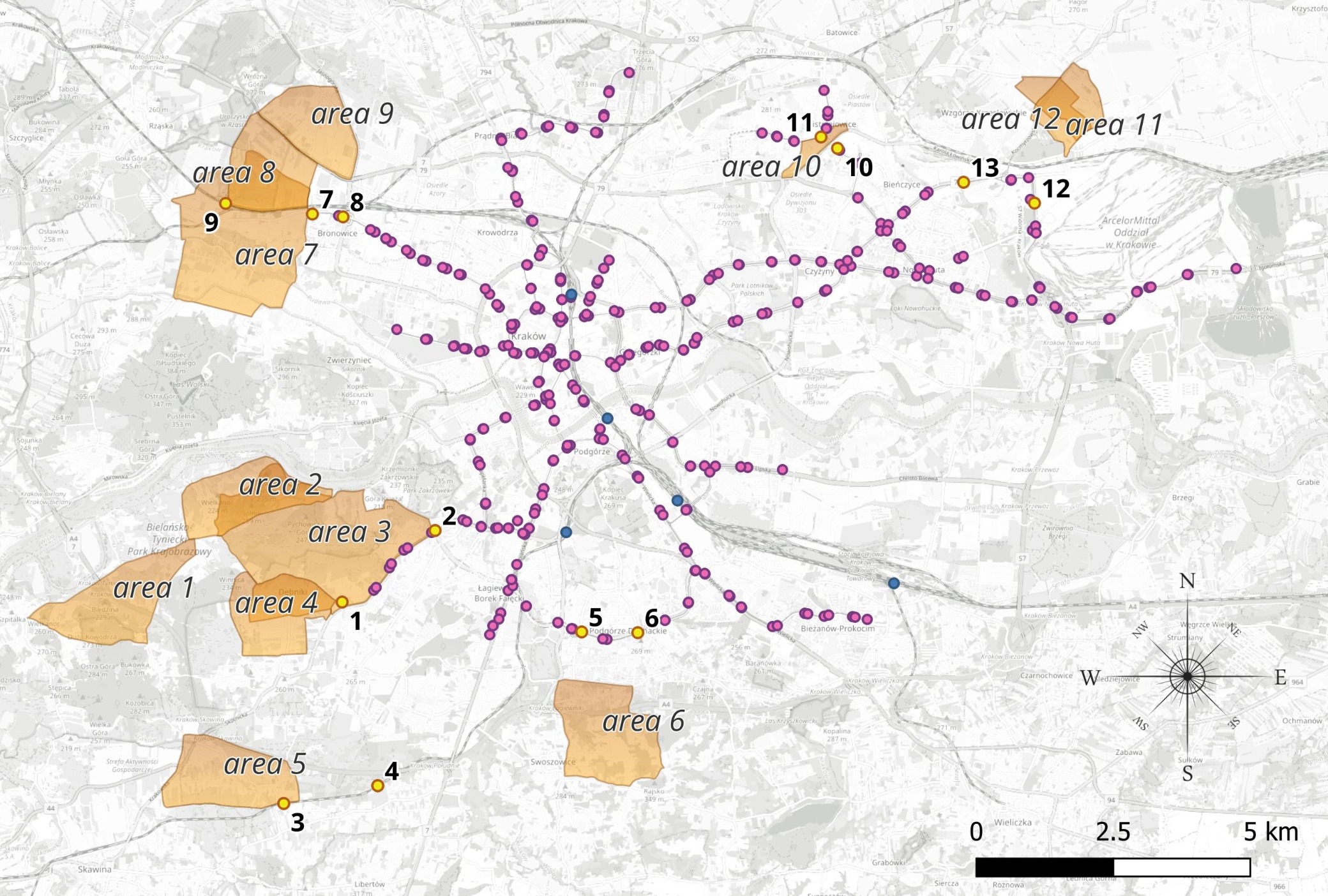}
	\caption{Krakow preselected areas (in orange) with corresponding hubs (in yellow), tram stops (in pink) and train stops (in blue).}
	\label{FIG:2}  
    \end{minipage}\hfill
    \begin{minipage}{0.45\textwidth}
        \centering
		\includegraphics[width=\linewidth]{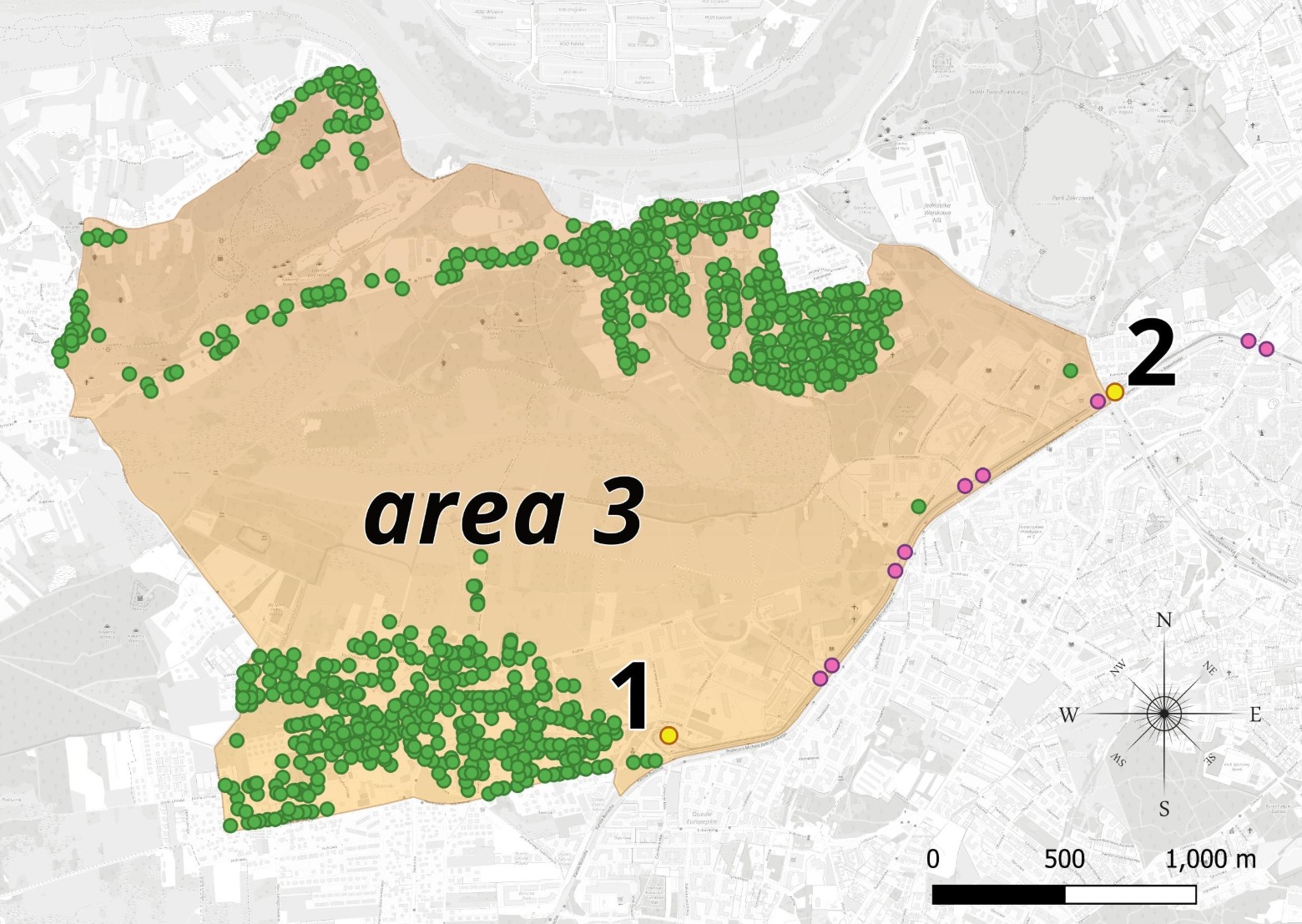}
	\caption{Example of the spatial distribution of address points (in green), tram stops (in pink) and light rail hubs (in yellow) for the Area~3.}
	\label{FIG:3}    
    \end{minipage}
\end{figure}

For example, Krakow, Poland, plans to launch an on-demand bus service integrated with PT in one of the 12 low-density area candidates (Fig.~\ref{FIG:2}). During morning rush hours, the operator will collect ride requests from travellers heading to hubs, dispatching small-capacity buses for pooled rides. These buses will pick up travellers from designated stops and transport them to high-frequency tram or train hubs (Fig.~\ref{FIG:3}), enabling further travel by efficient public transport. This feeder service aims to increase accessibility to PT and address first-mile challenges. However, a critical question remains: Where should the service be implemented, given the uncertainties about which travellers will use it.

\subsection{Literature review}


Demand-responsive transit (DRT) solutions, such as feeder systems, offer a flexible and adaptive alternative to traditional fixed-route transit services. These systems adapt to real-time passenger demand, making them particularly effective in low-demand areas where conventional public transportation often struggles with high operational costs and low ridership. By enabling ride requests via digital platforms, such as mobile apps, DRTs improve accessibility and user satisfaction by effectively addressing the connectivity challenges of the first and last mile. This provides seamless links to larger transit networks and improves the overall efficiency of mass transit systems \citep{leffler2021simulation}. Implementing a feeder system requires careful planning to address specific regional needs. Key factors include the type of connection (e.g., suburban-to-urban, as discussed in \citet{badia2021design} research, or urban-focused, as explored in \citet{shen2018integrating} research), route flexibility \citep{badia2021design,leffler2021simulation, li2010feeder}, automation and electrification \citep{badia2021design}, different demand patterns, many-to-one and many-to-many \citep{kuah1989feeder} etc. Ride-pooling, a subset of DRT that pools passengers travelling in the same direction into a single vehicle, further optimises service by reducing operational costs, travel times and environmental impact \citep{alonso2017demand}, making it particularly valuable in low-density areas where traditional buses may remain underutilised.

Assessing on-demand feeder services requires comprehensive KPIs to evaluate performance, efficiency, and user satisfaction. The choice of KPIs depends on the characteristics of the service. \citet{leffler2021simulation} evaluate station-based automated feeder services using passenger costs, travel time components, and total vehicle kilometres to analyse trade-offs between fixed and on-demand service designs (Stockholm case study). \citet{badia2021design} assess the impact of automation and electrification on fixed and door-to-door feeder services, focussing on cost structures. \citet{li2010feeder} propose metrics combining customer walking, waiting, and ride times to assess demand-responsive connector services, offering an alternative to fixed routes in specific conditions. \citet{shen2018integrating} explore autonomous vehicles for first-mile connections in Singapore, evaluating travel time, traffic impacts, and financial viability. \citet{calabro2022fixed} assess many-to-one demand patterns, analysing spatial distribution, service area, and configurations using total unit costs, including passenger delays and operator detours.
Accessibility is also a key factor in evaluating on-demand feeder services. Their effectiveness lies in connecting users to transit hubs efficiently \citep{chandra2013accessibility}. Research shows that proximity to transit stops (400–800 metres from homes) strongly influences public transport use  \citep{tennoy2022walking}. In areas with limited public transport networks, on-demand feeders can reduce travel distances to hubs, increasing public transport uptake and reducing car dependency. However, success depends on optimal stop placement and route planning to ensure user convenience.
\citet{shulika2024spatiotemporal} focuse on ride-pooling as an on-demand feeder service for many-to-many demand patterns. While their work uses six KPIs to balance traveller, driver, operator, and policymaker interests, we assess factors like vehicle mileage reduction, passenger comfort, and occupancy in simulations for many-to-one demand patterns typical of feeder services with a single transfer point.
The selection of areas for implementing on-demand feeders is complex. Many studies present them as effective alternatives to complement or replace fixed-route systems \citep{badia2021design, leffler2021simulation, li2010feeder}, or model scenarios in idealised square cities to optimise network design and service areas \citep{badia2021design}. 
Recent work has begun to address more systematically the question of where and under what conditions DRT performs best. \citet{wang2025planning} show that planning DRT with the conventional objective of minimising generalised cost, without explicit attention to the distribution of accessibility, can inadvertently exacerbate rather than reduce the accessibility gap between well-served corridors and underserved suburbs. They propose a bilevel optimisation strategy guided by a scoring function that targets areas combining high population density with low PT accessibility; tested on models of Montreal, Budapest and Lisbon, the method reduces accessibility inequality by more than $20\%$ as measured by the Atkinson index. \citet{calabro2025and} address the same question from an operational angle, using a parametric agent-based model to compare DRT performance across three representative contexts: many-to-many demand in small cities, off-peak demand in large cities, and many-to-one suburban feeder services. Their results show that performance is fundamentally context-dependent and highly sensitive to the fleet-to-demand ratio; suburban feeder applications emerge as the most cost-efficient setting, achieving total unit costs of 4–6~\euro/passenger with a fleet of only 4–6~vehicles representing $40-60\%$ cost advantages over urban scenarios. Both studies confirm that spatial and demand context shape DRT viability decisively, yet neither addresses the pre-deployment situation where demand data do not yet exist – which is precisely the gap our method fills.
\subsection{Methodological approach and contributions}

One of the key challenges in launching new on-demand feeder services in urban areas that lack high-frequency public transport is the absence of reliable demand data at the planning stage. While recent work has advanced our understanding of where DRT tends to perform well \citep{wang2025planning, calabro2025and}, these approaches rely on either observed demand distributions or synthetic demand assumptions – data that are rarely available to municipalities considering an entirely new service. Traditional methods often resort to parameter tuning, adjusting predefined service configurations against assumed demand levels. This introduces substantial uncertainty when actual ridership patterns are unknown.

We propose a parameter-free approach that utilises probabilistic demand fractions to simulate potential demand without requiring prior calibration of demand-specific parameters, such as observed trip rates, modal split, or route-level ridership, which are typically unavailable before a new service is launched. The ExMAS algorithm itself relies on behavioural cost parameters (value of time, discomfort penalty, and fare per kilometre) drawn from published literature and local data sources; these do not constitute demand-specific inputs and do not require local ridership observations. This approach enables municipalities to conduct a pre-deployment assessment of different urban areas, identifying those most suitable for on-demand feeder bus services even in the absence of precise demand data. We consider the fraction of residents who must be interested in the service sufficient to achieve minimum required efficiency thresholds. Areas requiring the lowest resident interest to meet these thresholds are considered the most favourable for service implementation.

To assess which area has the greatest potential before the service is introduced, we identify those that:
\begin{itemize}
    \item  require a lower level of demand to achieve efficiency;
    \item  reach stability at the lowest demand level in performance indicators, marking the required threshold;
    \item perform better at expected demand levels (if known).   
\end{itemize} 

After identifying the most promising area and hub combination, we establish benchmarks to understand the service's early-stage performance. 

\section{Method}

Our approach for selecting the preferred area to implement on-demand pooled transit feeders integrated with public transport (PT) is illustrated in Fig.~\ref{FIG:1}. The process begins with data collection, including population distribution, road network, candidate areas, and public transport hub locations (one or more per area).

\begin{figure}
	\centering
		\includegraphics[width=0.7\linewidth]{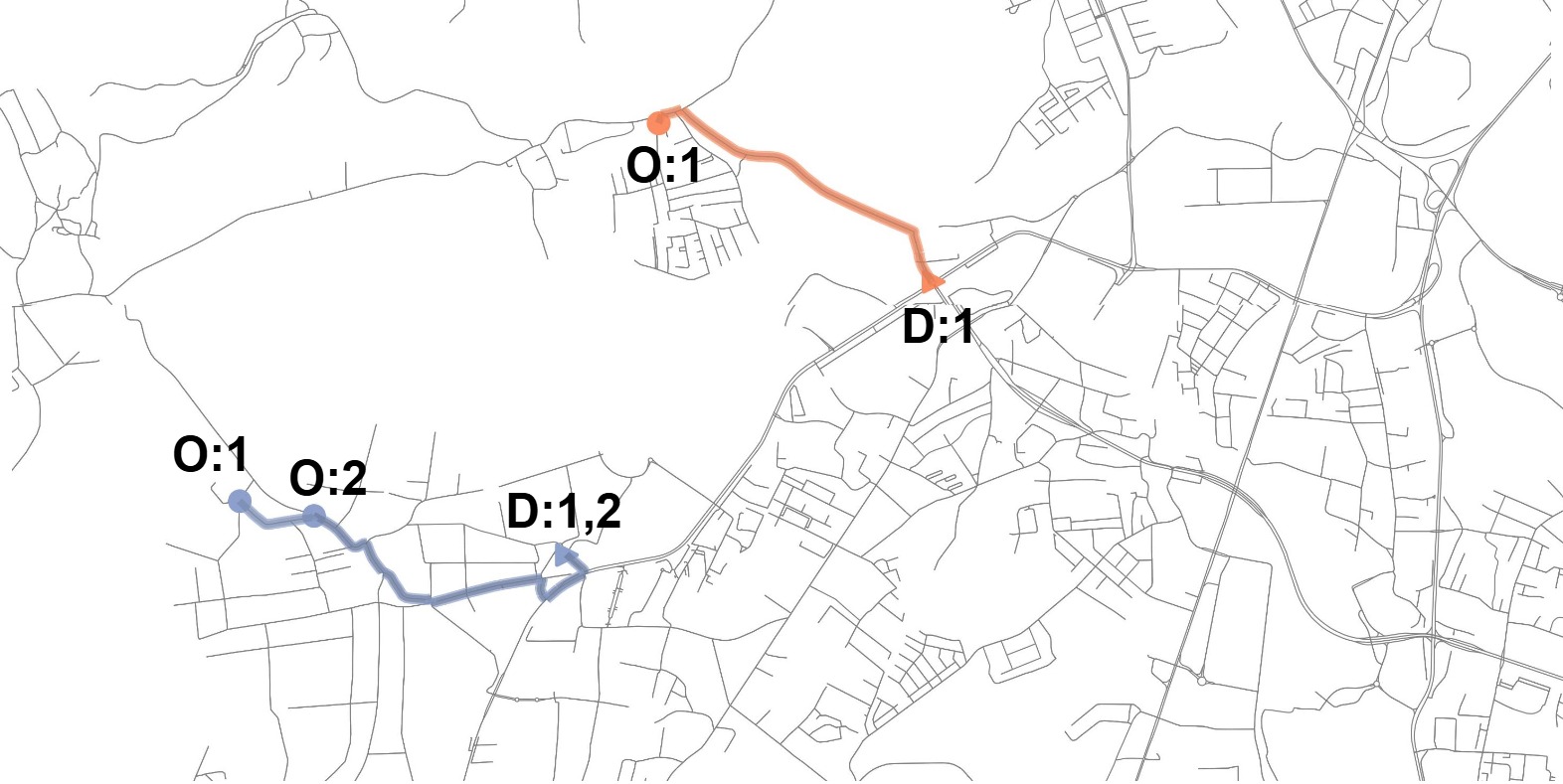}
	\caption{An example of a road network graph of Area 3 shows a shared ride (blue) by two travellers to hub 1, and a non-shared, private ride (orange) to hub 2. The dots represent the origins, and the triangles indicate transit destinations (hubs).}
	\label{FIG:4}    
\end{figure}

For each combination of area, hub location(s), demand fraction level, and replication, we apply the utility-based Exact Matching of Attractive Shared Rides (ExMAS) algorithm \citep{kucharski2020exact} to match travellers to pooled rides. We start with a scenario where all travellers use individual ride-hailing, then assess the extent to which ride-hailing can be replaced by ride-pooling while maintaining traveller satisfaction and vehicle mileage reduction. The ride-hailing baseline is a deliberate modelling choice: $\alpha $ captures the fraction of residents willing to use the on-demand service, and the KPIs quantify what pooling adds over serving those same residents individually. This keeps the comparison consistent across all candidate areas and separates the pooling gain from questions of modal competition, which are beyond the scope of a pre-deployment assessment.
Travellers who cannot be efficiently matched to shared rides continue using individual ride-hailing (Fig. \ref{FIG:4}).
Next, we evaluate the potential of on-demand service using three key indicators (based on \citet{shulika2024spatiotemporal}): mileage reduction, passenger satisfaction, and occupancy. We assess system performance using KPIs derived from the simulation results. 

We first track the progression of these KPIs across varying demand fractions to identify the most promising hub within each preselected area. We then compare candidate areas by identifying the fraction of residents (demand level, $\alpha$) required to meet the efficiency thresholds for launching the service. 
The three thresholds represent the minimum operational signal that a pooled service is generating any benefit relative to solo ride-hailing, rather than targets for a mature system. The vehicle-hours reduction threshold ($\Delta T_v \geq 0.1$, i.e. $10\%$) is informed by values reported in the ride-pooling literature for low-to-moderate demand conditions. \citep{kucharski2020exact} report that even modest pooling (with an average occupancy of around 1.67) can achieve vehicle-hours reductions of approximately $30\%$ in a dense urban setting. In the low-density, low-demand feeder context considered here, a $10\%$ reduction is therefore a realistic minimum signal that pooling is generating any operational gain at all. \citep{shulika2024spatiotemporal} similarly observe that mileage reductions in the range of $10-20\%$ are characteristic of the early, sub-critical demand phase before the system reaches critical mass. The passenger comfort threshold ($\Delta U_p \geq 0.025$, i.e. $2.5\%$ improvement in perceived utility) is intentionally set at a low level. In the scenario analysed here, a free on-demand feeder service – travellers do not pay a fare premium for pooling, so the utility trade-off is driven primarily by detour and waiting time. A $2.5\%$ improvement ensures that, on average, travellers are not perceptibly worse off by sharing, which is a necessary condition for voluntary uptake of the service. A more stringent comfort threshold would be appropriate for a fare-charging service, but is not warranted here. The occupancy threshold ($O \geq 2$) is the logical minimum for any pooling to occur at all: an average occupancy below 2 implies that the vast majority of trips remain solo rides, and the service is effectively operating as individual ride-hailing. This interpretation is consistent with the use of occupancy as a pooling indicator in the literature \citep{shulika2024spatiotemporal, kucharski2020exact}, and with regulatory definitions of shared mobility in several jurisdictions that require an average occupancy exceeding 2 to qualify as a pooled service. Together, the three thresholds define the lowest fraction of demand at which a feeder service functions as a pooled system rather than as a collection of solo trips. Areas that meet all three thresholds at low demand fractions are therefore the most promising candidates for early-stage deployment.

Therefore, in this study, we have set the following efficiency thresholds for launching the service:

\begin{itemize}
    \item  $\triangle{T}_{v} (\text{vehicle hours reduction})\geq 0.1$: the launching of shared rides (instead of individual ones) allows for a reduction of vehicle kilometres by at least $10\%$;
    \item $\triangle{U}_{p} (\text{travellers utility gains})\geq 0.025$: passenger comfort improves by at least $2.5\%$  compared to individual travel. For the analysed scenario of free on-demand bus service, this measure ensures that passengers do not encounter significant discomfort associated with a new service;
    \item ${O} (\text{occupancy}) \geq 2$: the average vehicle occupancy exceeds 2.   

\end{itemize}

After selecting the most promising area-hub combination, we establish benchmarks to provide municipalities with insights into the new service's potential, including the minimum demand fraction needed to pool travellers, the demand level where ride-pooling potential grows, and the demand level required for consistent KPI achievement.
Scripts for reproducible results are available in the public repository. 

\subsection{Input}

The presented method relies on several key inputs: OpenStreetMap (OSM) data detailing the city's road network; the distribution of the city's population, including the coordinates of residential address points; and a set of candidate areas along with their locations near the public transport hub for a comparative evaluation.

The candidate areas and their associated public transport hubs are provided by the analyst or the relevant municipality planning authority. An area may be associated with more than one hub where geographic proximity or public transport network connectivity makes multiple hubs plausible destinations for feeder trips. In such cases, demand is generated separately for each hub, and the hub delivering the best KPI performance is selected as the preferred hub ${H}^{*}$ for that area (see Section 2.6).

The proposed method is nondeterministic because for a given $\alpha$ the number of travellers is defined, but each of them is randomly selected from the entire population; therefore, the results vary between replications. To obtain meaningful results, we replicate the process and draw conclusions from both individual realisations and aggregated values of three KPIs.

\subsection{Demand generation}

\begin{figure}
	\centering
		\includegraphics[width=\linewidth]{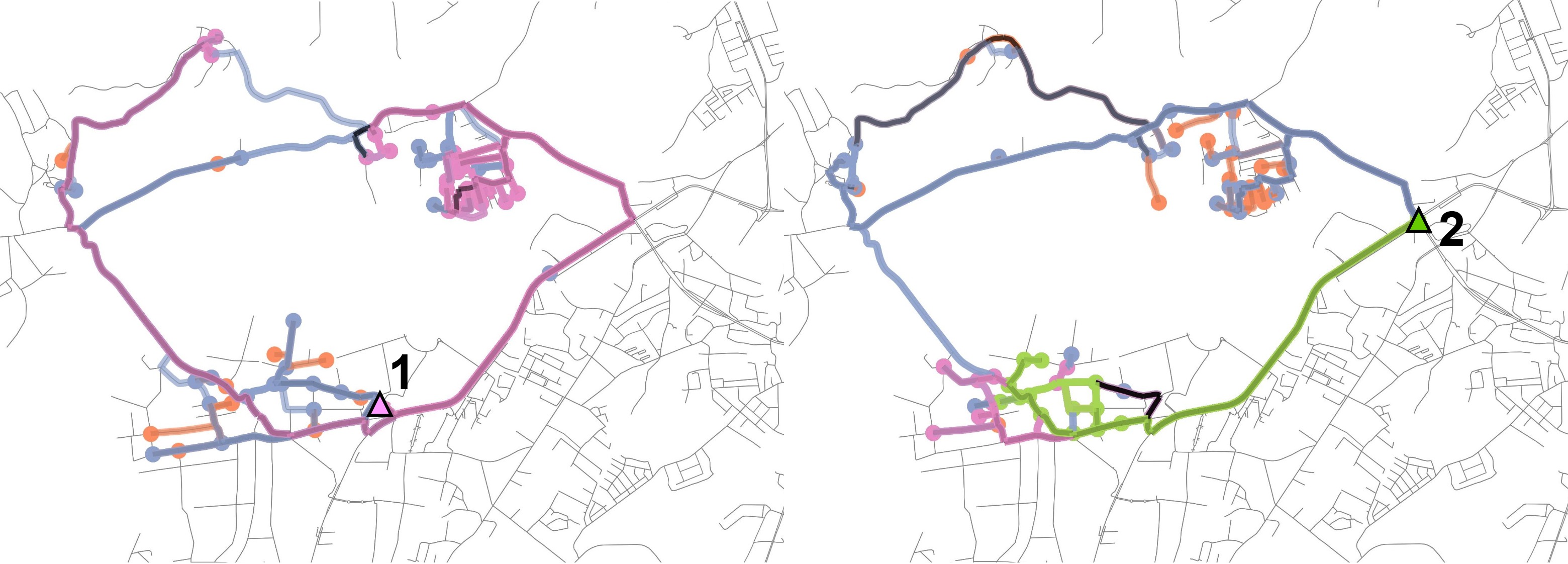}
	\caption{An example of visualising ride-pooling algorithm shows sample rides for Area 3: all sampled travellers of an area are heading from origins (dots) to hub 1 (left pink triangle denoted 1) or to hub 2 (right green triangle denoted 2) as their transit destination points (hubs).}
	\label{FIG:5}    
\end{figure}

For each candidate area $A$, we determine a total resident population based on demographic data, which represents the area’s maximum potential demand. We then evaluate the potential of ride-pooling at various demand fractions, up to $5\%$ of the area population, covering a range from the minimum necessary for pooling feasibility to levels sufficient for stable KPI results. We assume that demand is generated randomly, reflecting the population distribution within each area.  Starting with a $0.1\%$ demand fraction for all preselected areas $A$, we gradually increase it to $5\%$, selecting a sample of travellers with residential address points as origins $\left\{o_i\right\}$. All travellers are assumed to head to the nearest public transport hub $H$ in their area as their destination $d$ (Fig. \ref{FIG:5}). 
In cases where an area has multiple hubs, demand is generated separately for each hub, first using one as the destination and then repeating for the others (Fig.~\ref{FIG:5}). Each travel request time $\tau_i$ is sampled uniformly within the simulation period. This sampling approach is chosen because, in the absence of observed arrival data, any assumed temporal pattern could bias the analysis. The uniform distribution represents the most neutral assumption, making no claim about within-window peaking behaviour. While low-demand areas may exhibit substantial variability in request timing, representing such dynamics would require observed travel demand data that are not available in the present study. Replacing it with an empirical arrival distribution is a natural next step once the monitoring data exist. To conclude, we generate a set of travel requests $\left\{q_i=\left (o_i, d, \tau_i\right ) \right\}$ for each area-hub pair, each demand fraction and replication.

\subsection{ExMAS}
ExMAS is an open source algorithm, implemented using the Python programming language, that solves the offline ride-pooling problem for travellers \citep{kucharski2020exact}. For a potential demand set of trips (based on the demand fraction levels identified above), we use ExMAS to identify all feasible and attractive door-to-door pooled rides (groups of travellers willing to share a trip). Then the algorithm optimally assigns each traveller to a vehicle in a way that minimises total mileage. ExMAS is utility-based and assumes that pooling is selected only if it is attractive, i.e., travellers opt for shared trips only if the utility of pooling exceeds the utility of solo travel, based on travel time (adjusted by the value of time) and cost (subject to fare per kilometre and distance). 
ExMAS identifies all attractive shared rides by considering the trade-off between costs (detours, delays, discomfort) and benefits (reduced fares). It assumes that travellers choose shared rides only if they are more appealing than solo rides. ExMAS solves the matching problem to minimise total mileage while ensuring that each traveller is uniquely assigned to a ride. The solution may include shared or solo rides. ExMAS is exact, explores all ride combinations, and aligns with discrete choice theory. However, it requires a priori known demand and does not explicitly model the vehicle fleet. 
The full utility formulation, including the treatment of detour, pick-up delay, discomfort multiplier, and fare discount, is detailed in \citet{kucharski2020exact} and \citet{shulika2024spatiotemporal}; the behavioural parameter values used in the present study are reported in Section~3.1. Since ExMAS does not explicitly model fleet size, the reported KPI values represent an upper bound on pooling potential. In practice, a limited fleet would render some matched rides infeasible when simultaneous demand exceeds vehicle availability, reducing effective pooling rates and occupancy. Acceptable waiting time is governed endogenously by the traveller utility function rather than by a hard constraint: a traveller accepts a pooled ride only if the utility gain from the fare discount outweighs the discomfort of detour and pick-up delay, so the implicit waiting time tolerance varies with trip length and behavioural parameters. In the low-demand scenarios central to this study, fleet constraints are unlikely to be binding; explicit fleet modelling is identified as a direction for future work (Section~4.1).
 In the scenario analysed here, the feeder service is modelled as free at the point of use for passengers, meaning that the fare trade-off does not drive pooling decisions; instead, travellers accept shared rides when the utility loss from detour and pick-up delay (weighted by the discomfort multiplier $\beta_s = 1.3$) is sufficiently compensated. In paid-service contexts, fare elasticity is a well-documented driver of pooling uptake \citep{de2023ride}, and incorporating a probabilistic treatment of behavioural parameters following the approach of \citet{bujak2024ride} is identified as a direction for future research.

\subsection{Performance indicators}

We report three KPIs of interest (for full definitions of possible KPIs we refer to \citet{shulika2024spatiotemporal}). For each combination of area $A$, hub $H$ and a single realisation of demand with given fraction of demand $\alpha$, we report: 

\begin{itemize}
    \item  what is the potential mileage reduction $\Delta T_v(\alpha, A, H)$;
    \item  how is the (perceived) traveller utility improved $\Delta U_p(\alpha, A, H)$;
    \item  what is the average occupancy  $O(\alpha, A, H)$.
\end{itemize} 

 Potential mileage reduction and passenger comfort are defined by comparing vehicle hours and traveller utility when the ride-pooling service is available and when it is not applied. Occupancy represents the ratio of total passenger hours in the solo ride-hailing scenario to total vehicle hours in the pooled scenario \citep{shulika2024spatiotemporal}.

\subsection{Thresholds}

We collect the result from the broad range of $\alpha$'s for each area $A$ and hub $H$, replicable enough to draw conclusions. To make comparisons meaningful, we set a threshold for each of the above KPIs and answer the reverse question: What is the minimal $\alpha^*$ such that a given threshold is met?

The thresholds are as follows:

\begin{subequations}\label{eq:thresholds_all}
    \begin{align}
        \alpha_{\Delta T_v}^*(A,H) &= \underset{\alpha \in [0.001, 0.05]}{\mathrm{min}}\Delta T_v(\alpha, A, H) \geq 0.1, \label{eq:threshold_time} \\ 
        \alpha_{\Delta U_p}^*(A,H) &= \underset{\alpha \in [0.001, 0.05]}{\mathrm{min}}\Delta U_p(\alpha, A, H) \geq 0.025, \label{eq:threshold_utility} \\ 
        \alpha_{O}^*(A,H) &= \underset{\alpha \in [0.001, 0.05]}{\mathrm{min}}O(\alpha, A, H) \geq 2.0. \label{eq:threshold_occupancy} 
    \end{align}
\end{subequations}

We are interested in areas that surpass the required levels of performance with the lowest possible demand level~$\alpha$.

\subsection{Hub and area selection}

To select the most promising combination of area and hub location for each area, we first identify which hub $H$ for each area  $A$ holds the most potential, denoted as ${H}^{*}$. 
The hub that achieves the best performance becomes the optimal hub for that area.

Then we determine at which demand levels $\alpha$ the combination of area $A$ and hub ${H}^{*}$ meets KPIs thresholds.
The area $A$ and hub ${H}^{*}$ that meets these thresholds with the lowest demand fraction $\alpha$ becomes the most potential area, denoted as ${A}^{*}$.

 \subsection{Benchmarks}
After selecting the most promising combination (${A}^{*}, {H}^{*}$) we run ExMAS simulations across demand fractions $\alpha\in[0,0.01]$ in 30 replications. This allows us to explore the most unstable phase, which typically occurs during the initial launch when demand is at its lowest. We establish the following benchmarks:

 \begin{itemize}
    \item minimum pooling demand: the minimum fraction of demand $\alpha$ required to successfully pool travellers into shared trips;
    \item KPI growth point: the demand fraction $\alpha$ at which ride-pooling potential starts to grow;
    \item  consistent KPI achievement: the demand fraction $\alpha$ at which a stable level of all three KPIs is reached.     
\end{itemize} 

These benchmarks help set realistic expectations for performance in a specific early stage of service.

\subsection{Summary}

The proposed methodology allows identifying the most promising areas and public transport hub to launch on-demand pooled transit feeder services when the exact demand remains unknown. We specifically focus on the critical early phase of launching a new service, characterised by low demand. We examine how the system performs under varying demand levels. 

By replicating the assessment for different demand patterns, we obtain more reliable results.
These findings will guide the decision-making process in selecting the optimal locations and conditions to deploy on-demand pooled transit feeder services within the urban transport network.

\section{Results}
\subsection{Experimental settings}

In this experiment, we use our method to compare the ride-pooling potential in 12 preselected areas of Krakow (Poland), each paired with the nearby public transport hub(s). The input data used in the analysis is the Krakow road network (sourced from OSM), the population distribution that details the residents within each preselected area (provided by the municipality) and the locations of public transport hubs throughout Krakow. The objective is to explore how well candidate areas perform when implementing on-demand pooled transit feeders, focussing on three KPIs: vehicle kilometre reduction ($\Delta T_v$), passenger comfort ($\Delta U_p$), and vehicle occupancy ($O$). Since the exact demand is unknown, the performance of these KPIs is assessed across varying demand fractions $\alpha$, representing the proportion of the area’s population that could use the service.

Table \ref{tab1} shows 12~candidate areas in Krakow, along with their corresponding public transport hubs. The numbers of areas and hubs correspond to Fig. \ref{FIG:2}. 
Areas vary significantly in terms of population density, with Area 10 having the highest density at 9880.8 residents per square kilometre, while Area 1 has the lowest density at 489.8 residents per square kilometre. In terms of proximity to public transport hubs, the average distance to the nearest hub ranges from $0.374~km$ in Area~10 to $6.183~km$ in Area 1, with shorter distances potentially reducing the need for feeder services. Some areas also have access to multiple public transport hubs, which can improve accessibility and reduce the need for feeder services. Furthermore, we report the type of public transport hub, such as tram stops (e.g., Hub 1. ’Czerwone Maki P+R’) or train stations (e.g., Hub 3. ’Krakow Sidzina’).

\setlength{\aboverulesep}{0.4pt}
\setlength{\belowrulesep}{0pt}
\begin{table}[ht]
\caption{Characteristics of preselected areas}\label{tab1}
\renewcommand{\arraystretch}{1.1}
\setlength{\tabcolsep}{3pt} 
\begin{tabularx}{\textwidth}{
    >{\centering\arraybackslash}p{0.9 cm}
    >{\centering\arraybackslash}p{0.8 cm}
    >{\centering\arraybackslash}p{1.3 cm}
    >{\centering\arraybackslash}p{2 cm}
    >{\centering\arraybackslash}p{3.7 cm}|
    >{\centering\arraybackslash}p{1 cm}|
    >{\centering\arraybackslash}p{1.6 cm}}
\toprule%
\multirow{2}{*}{Name}& \multirow{2}{*}{Surface} & Population & Density & \multicolumn{3}{@{}c@{}}{Hub} \\\cmidrule{5-7}%
&  & [residents] & [residents/km\textsuperscript{2}] & Name & Type & Avg. distance to hub [$km$] \\
\midrule
    Area 1  & {\includegraphics[width=8mm,height=6mm] {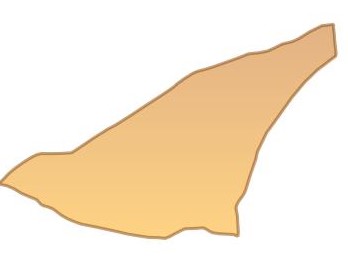}} & 1263 & 489.8 & {1.'Czerwone Maki P+R'} & \raisebox{0pt}{\includegraphics[width=8mm,height=4mm] {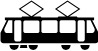}}& 6.183\\
\midrule
    \multicolumn{1}{c}{\multirow{2}{*}{\begin{tabular}[c]{@{}c@{}}Area 2\end{tabular}}} & \multirow{2}{*}{{\includegraphics[width=8mm,height=6mm] {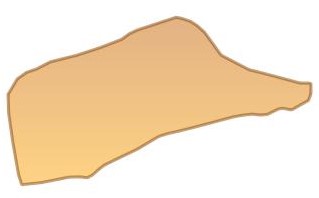}}} & \multirow{2}{*}{1286} & \multirow{2}{*}{585.6} & {1.'Czerwone Maki P+R'} & {\includegraphics[width=8mm,height=4mm] {Figs/Tram.jpg}} & 3.385 \\ & & & & {2.'Norymberska'} & {\includegraphics[width=8mm,height=4mm] {Figs/Tram.jpg}} & 5.416\\
\midrule
        \multicolumn{1}{c}{\multirow{2}{*}{\begin{tabular}[c]{@{}c@{}}Area 3\end{tabular}}} & \multirow{2}{*}{{\includegraphics[width=8mm,height=6mm] {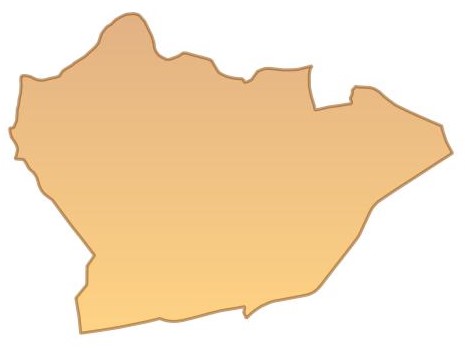}}} & \multirow{2}{*}{4550} & \multirow{2}{*}{676.9} & {1.'Czerwone Maki P+R'} & {\includegraphics[width=8mm,height=4mm] {Figs/Tram.jpg}} & 1.762 \\ & & & & {2.'Norymberska'} & {\includegraphics[width=8mm,height=4mm] {Figs/Tram.jpg}} & 3.454\\
\midrule
        Area 4  & {\includegraphics[width=8mm,height=6mm] {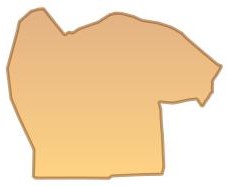}} &3719 & 1781.2 & {1.'Czerwone Maki P+R'} & {\includegraphics[width=8mm,height=4mm] {Figs/Tram.jpg}}& 1.744\\
\midrule
        \multicolumn{1}{c}{\multirow{3}{*}{\begin{tabular}[c]{@{}c@{}}Area 5\end{tabular}}} & \multirow{3}{*}{{\includegraphics[width=8mm,height=6mm] {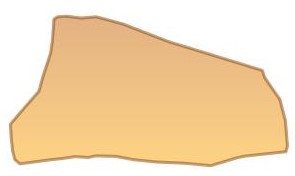}}} & \multirow{3}{*}{1593} & \multirow{3}{*}{658.6} & {1.'Czerwone Maki P+R'} & {\includegraphics[width=8mm,height=4mm] {Figs/Tram.jpg}} & 4.369 \\ & & & & {3.'Krakow Sidzina'} & {\includegraphics[width=4mm,height=4mm] {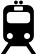}} & {1.634}\\& & & & {4.'Krakow Opatkowice'} & \raisebox{0pt}{\includegraphics[width=4mm,height=4mm] {Figs/Train.jpg}} & 4.163\\
\midrule
        \multicolumn{1}{c}{\multirow{2}{*}{\begin{tabular}[c]{@{}c@{}}Area 6\end{tabular}}} & \multirow{2}{*}{{\includegraphics[width=8mm,height=6mm] {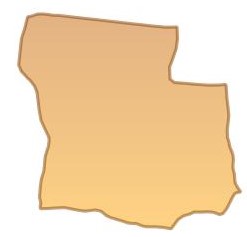}}} & \multirow{2}{*}{2396} & \multirow{2}{*}{851} & {5.'Kurdwanów P+R'} & {\includegraphics[width=8mm,height=4mm] {Figs/Tram.jpg}} & 1.762 \\ & & & & {6.'Nowosadecka'} & {\includegraphics[width=8mm,height=4mm] {Figs/Tram.jpg}} & 2.077\\
\midrule
        \multicolumn{1}{c}{\multirow{2}{*}{\begin{tabular}[c]{@{}c@{}}Area 7\end{tabular}}} & \multirow{2}{*}{{\includegraphics[width=8mm,height=6mm] {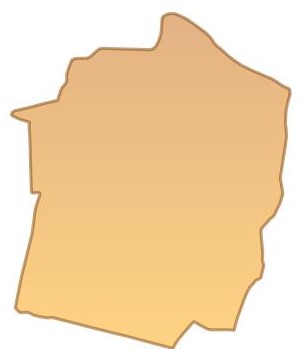}}} & \multirow{2}{*}{5651} & \multirow{2}{*}{676.9} & {7.'Bronowice Małe'} & {\includegraphics[width=8mm,height=4mm] {Figs/Tram.jpg}} & 2.309 \\ & & & & {9.'Kraków Mydlniki(PKP)'} & {\includegraphics[width=4mm,height=4mm] {Figs/Train.jpg}} & 1.553\\
\midrule
        \multicolumn{1}{c}{\multirow{2}{*}{\begin{tabular}[c]{@{}c@{}}Area 8\end{tabular}}} & \multirow{2}{*}{{\includegraphics[width=8mm,height=6mm] {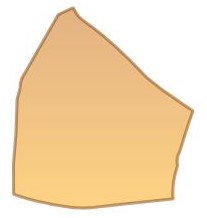}}} & \multirow{2}{*}{1836} & \multirow{2}{*}{904.5} & {7.'Bronowice Małe'} & {\includegraphics[width=8mm,height=4mm] {Figs/Tram.jpg}} & 1.072 \\ & & & & {9.'Kraków Mydlniki(PKP)'} & {\includegraphics[width=4mm,height=4mm] {Figs/Train.jpg}} & 1.924\\
\midrule
        \multicolumn{1}{c}{\multirow{3}{*}{\begin{tabular}[c]{@{}c@{}}Area 9\end{tabular}}} & \multirow{3}{*}{{\includegraphics[width=8mm,height=6mm] {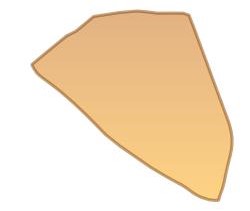}}} & \multirow{3}{*}{4002} & \multirow{3}{*}{2202.2} & {7.'Bronowice Małe'} & {\includegraphics[width=8mm,height=4mm] {Figs/Tram.jpg}} & 1.662 \\ & & & & {8.'Bronowice SKA'} & {\includegraphics[width=8mm,height=4mm] {Figs/Tram.jpg}} & {1.863}\\& & & & {9.'Kraków Mydlniki(PKP)'} & {\includegraphics[width=4mm,height=4mm] {Figs/Train.jpg}} & 2.911\\
\midrule
        \multicolumn{1}{c}{\multirow{2}{*}{\begin{tabular}[c]{@{}c@{}}Area 10\end{tabular}}} & \multirow{2}{*}{{\includegraphics[width=8mm,height=6mm] {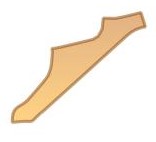}}} & \multirow{2}{*}{3069} & \multirow{2}{*}{9880.8} & {10.'Dunikowskiego'} & {\includegraphics[width=8mm,height=4mm] {Figs/Tram.jpg}} & 0.503 \\ & & & & {9.'Kraków Mydlniki(PKP)'} & {\includegraphics[width=8mm,height=4mm] {Figs/Tram.jpg}} & {0.374}\\
\midrule
        \multicolumn{1}{c}{\multirow{2}{*}{\begin{tabular}[c]{@{}c@{}}Area 11\end{tabular}}} & \multirow{2}{*}{{\includegraphics[width=8mm,height=6mm] {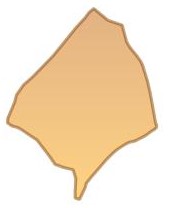}}} & \multirow{2}{*}{1925} & \multirow{2}{*}{1586.4} & {13.'Wańkowicza'} &{\includegraphics[width=8mm,height=4mm] {Figs/Tram.jpg}} & 3.243 \\ & & & & {12.'Zajezdnia Nowa Huta'} & {\includegraphics[width=8mm,height=4mm] {Figs/Tram.jpg}} & 1.916\\
\midrule
        \multicolumn{1}{c}{\multirow{2}{*}{\begin{tabular}[c]{@{}c@{}}Area 12\end{tabular}}} & \multirow{2}{*}{{\includegraphics[width=8mm,height=6mm] {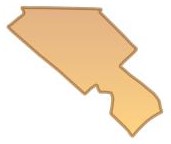}}} & \multirow{2}{*}{941} & \multirow{2}{*}{1334.3} & {13.'Wańkowicza'} & {\includegraphics[width=8mm,height=4mm] {Figs/Tram.jpg}} & 3.084 \\ & & & & {12.'Zajezdnia Nowa Huta'} & {\includegraphics[width=8mm,height=4mm] {Figs/Tram.jpg}} & 1.879\\

\bottomrule
\end{tabularx}
\end{table}

We set specific parameters for the simulation. To assess demand fractions, we define $\alpha$ values in two ranges. For each area-hub pair, demand fractions include $\alpha \in \{0.001, 0.002, 0.003, 0.005, 0.007, 0.009, 0.01, 0.02, 0.03, 0.05\}$. For the determination of benchmarks, we expand this range to include $\alpha \in \{0, 0.0002, 0.00025, 0.0004, 0.0005, 0.0006, 0.0007, 0.0008, 0.0009, 0.001, 0.0015, 0.002,$\\$0.003, 0.005, 0.006, 0.007, 0.008,  0.009, 0.01\}$. To ensure statistical reliability, demand generation is replicated multiple times: 10 replications for each area-hub pair and 30 replications for the most promising area-hub pair at each demand fraction. We also assume a vehicle capacity limit of six passengers, allowing six travellers to be accommodated in a single vehicle.

The ExMAS is parametrised with traveller behaviour: value-of-time ($\beta_t$, which we assume to be 43,3~$\text{PLN}/h$); penalty for sharing ( $\beta_s$, travel time multiplier due to discomfort of pooling, which we assume to be 1.3 based on the results of \cite{alonso2020value}); a trip fare $\beta_c$ of 6~$PLN/km$ (according to \cite{uber2022}); constant and network-wide flat speed of 21.6 $km/h$; each pick-up / drop-off operation takes 15 s. Simulation time: 30 minutes morning peak hour (7:45-8:15). We choose the 30-minute window to capture the core of the morning commute peak, when demand for feeder connections to PT hubs is most concentrated and pooling conditions are most favourable. Restricting the simulation to this window is also consistent with the uniform arrival assumption in Section 2.2: within a short, high-density interval, the uniform distribution serves as a more reliable approximation than it would be over a longer time window covering both peak and off-peak demand. This would shift the analysis away from the operational scenario for which the feeder service is intended.

\subsection{Hub selection}

\begin{figure}[!ht]
	\centering
		\includegraphics[width=\linewidth]{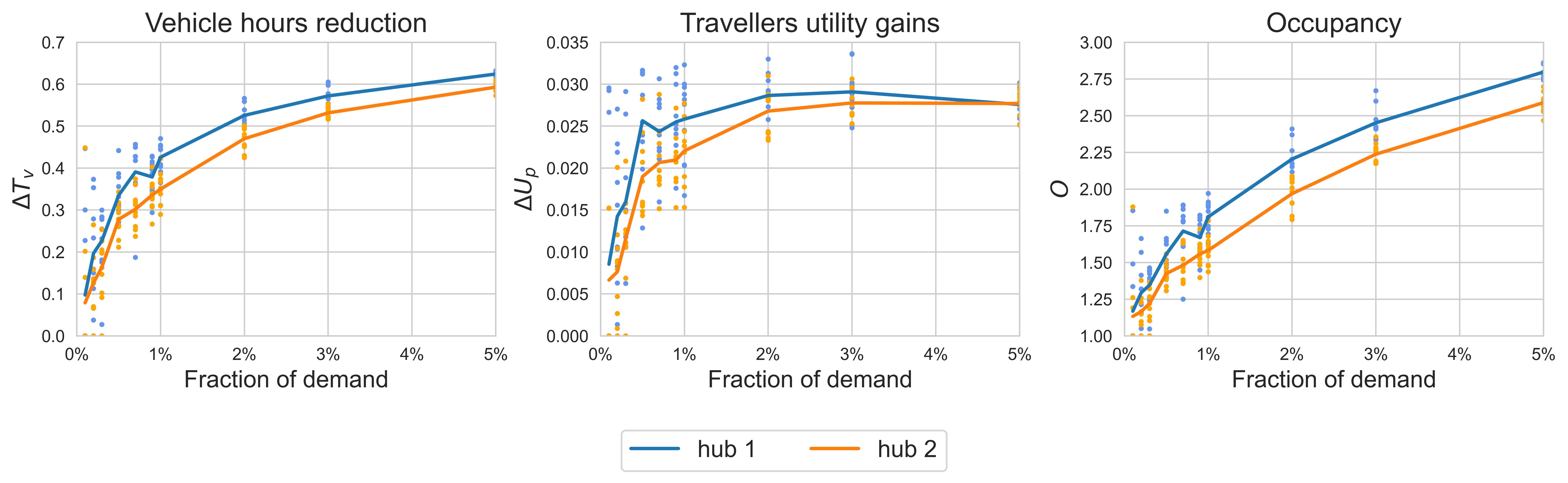}
	\caption{Three key performance indicators of ride-pooling plotted against the fraction of demand for Area 3. The lines represent the average performance across multiple simulations, while the dots represent individual simulation results. Both hubs in Area~3 showed similar trends, but hub 1 has a slight edge in potential.}
	\label{FIG:6}    
\end{figure}

The proposed method effectively selects the preferred hub location within each area and identifies the most promising areas in general to implement pooled transit feeders. Fig.~\ref{FIG:6} illustrates the hub selection process using our approach, with Area 3 as an example. 
Each dot on the graphs represents an individual simulation result for a specific KPI in a single replication at a given demand level. The lines represent the average performance across multiple replications. Across three indicators, the average performance lines for hub~1 are generally higher than those for hub~2, indicating that hub~1 demonstrates greater potential within the given demand range. Consequently, hub 1 is identified as the preferred hub for Area 3. Moving forward, in comparisons between areas, Area~3 will be evaluated in combination with its preferred hub, hub~1.
Similarly, we define the preferred hubs for the remaining areas. The preferred hubs selected for each candidate area are presented in Table~\ref{tab2} under the column labelled 'Hub'.

\subsection{Preferred area}

\begin{figure}[!ht]
	\centering
		\includegraphics[width=\linewidth]{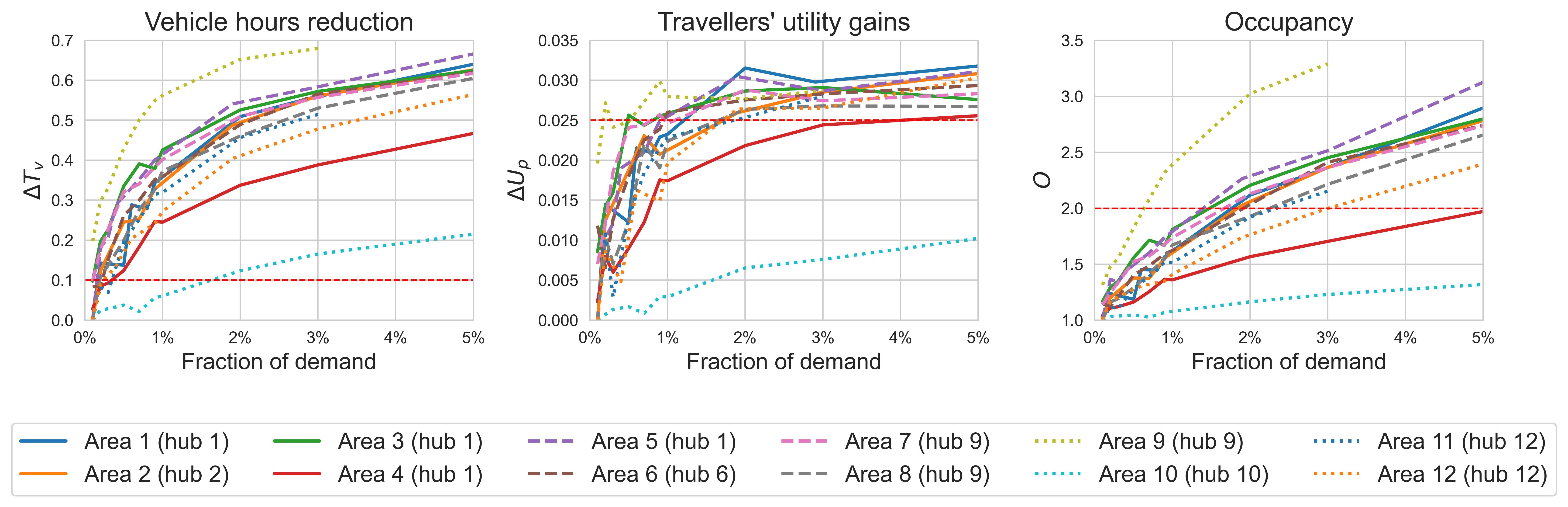}
	\caption{Three key performance indicators of ride-pooling plotted against the fraction of demand for preselected areas and the most promising corresponding hubs. Horizontal dashed red lines represent the set thresholds.}
	\label{FIG:7}    
\end{figure}

Fig. \ref{FIG:7}~illustrates the ride-pooling potential across preselected areas and their preferred hubs. Each dot represents a simulation result for a specific KPI at a given demand level, while the lines show average performance across multiple replications. Horizontal dashed lines indicate thresholds.
A critical aspect of this approach is to determine the minimum fraction of demand $\alpha$ that guarantees the achievement of each of the three thresholds, illustrated in Fig.~\ref{FIG:7} as horizontal dashed red lines. Table~\ref{tab2} provides the numerical values for this minimum demand level across all candidate area and preferred hub combinations, summarising the overall rank.

\begin{table}[h]
\caption{Characteristics of preselected areas}\label{tab2}
\renewcommand{\arraystretch}{1.2}
\setlength{\tabcolsep}{3pt} 
\begin{tabularx}{\textwidth}{
    >{\centering\arraybackslash}p{0.5cm}
    >{\centering\arraybackslash}p{3.7cm}
    >{\centering\arraybackslash}p{0.8cm} |
    >{\centering\arraybackslash}p{0.6cm} |
    >{\centering\arraybackslash}p{0.7cm} |
    >{\centering\arraybackslash}p{0.6cm} |
    >{\centering\arraybackslash}p{0.6 cm} |
    >{\centering\arraybackslash}p{0.5cm}
    >{\centering\arraybackslash}p{1.4cm}
    >{\centering\arraybackslash}p{1.0 cm}}
\toprule%
\multirow{3}{*}{Area}& \multirow{3}{*}{Hub} & \multicolumn{6}{c}{Threshold} &
      {Total} & {Final}\\\cmidrule{3-8}%
        &  &  \multicolumn{2}{c}{\textbf{$\Delta{T}_{v}\geq0.1$}} & \multicolumn{2}{|c|}{$\Delta{U}_{p}\geq 0.025$}& \multicolumn{2}{c}{${O}\geq2$} & {KPI rank} & {area} \\\cmidrule{3-8}%
        & & $\alpha$ & {rank} & $\alpha$ & {rank} & $\alpha$ & {rank} &  {score} & {rank} \\
\midrule
        \multicolumn{1}{c}{\begin{tabular}[c]{@{}c@{}} 1\end{tabular}} & {1.'Czerwone Maki P+R'} & 0.005 & 11 & 0.01 & 4 & 0.02 & 4 & 19 & 9 \\
        \multicolumn{1}{c}{\begin{tabular}[c]{@{}c@{}} 2\end{tabular}} & {2.'Norymberska'} & 0.002 & 4  & 0.02 & 7 & 0.02 & 4 & 15&6 \\
        \multicolumn{1}{c}{\begin{tabular}[c]{@{}c@{}} 3\end{tabular}} & {1.'Czerwone Maki P+R'} & 0.001 & 1  & 0.009 & 2 & 0.01 & 2 & 5&2 \\
        \multicolumn{1}{c}{\begin{tabular}[c]{@{}c@{}} 4\end{tabular}} & {1.'Czerwone Maki P+R'} & 0.003 & 10  & 0.05 & 11 & {-} & 11 & 32&11 \\ 
        \multicolumn{1}{c}{\begin{tabular}[c]{@{}c@{}} 5\end{tabular}} & {1.'Czerwone Maki P+R'} & 0.002 & 4  & 0.009 & 2 & 0.01 & 2 & 8&3 \\
        \multicolumn{1}{c}{\begin{tabular}[c]{@{}c@{}} 6\end{tabular}} & {6.'Nowosadecka'} & 0.002 & 4  & 0.01 & 4 & 0.02 & 4 & 12&5 \\
        \multicolumn{1}{c}{\begin{tabular}[c]{@{}c@{}} 7\end{tabular}} & {9.'Kraków Mydlniki (PKP)'} & 0.001 & 1  & 0.01 & 4 & 0.02 & 4 & 9&4 \\
        \multicolumn{1}{c}{\begin{tabular}[c]{@{}c@{}} 8\end{tabular}} & {9.'Kraków Mydlniki (PKP)'} & 0.002 & 4 & 0.02 & 7 & 0.02 & 4 & 15&6 \\
        \multicolumn{1}{c}{\begin{tabular}[c]{@{}c@{}} 9\end{tabular}} & {9.'Kraków Mydlniki (PKP)'} & 0.001 & 1  & 0.005 & 1 & 0.007 & 1 & 3&1 \\
        \multicolumn{1}{c}{\begin{tabular}[c]{@{}c@{}} 10\end{tabular}} & {10.'Dunikowskiego'} & 0.02 & 12  & {-} & 12 & {-} & 11 & 35&12 \\
        \multicolumn{1}{c}{\begin{tabular}[c]{@{}c@{}} 11\end{tabular}} & {12.'Zajezdnia Nowa Huta'} & 0.002 & 4 & 0.02 & 7 & 0.02 & 4 & 15&6 \\
        \multicolumn{1}{c}{\begin{tabular}[c]{@{}c@{}} 12\end{tabular}} & {12.'Zajezdnia Nowa Huta'} & 0.002 & 4 & 0.02 & 7 & 0.03 & 10 & 21 & 10 \\       
\bottomrule
\end{tabularx}
\end{table}

Areas that fail to meet any of the thresholds within the demand up to $5\%$ are indicated with a '-'. 
For each threshold, we rank the areas from 1 to 12, with 1 being the highest and 12 being the lowest. The area with the lowest demand fraction required to meet a threshold receives the highest rank for that threshold. This ranking process is repeated for all three thresholds, and the total KPI rank score is calculated for each area by summing the KPI ranks. Then we rank the areas, with lower totals indicating greater potential for successful implementation of the on-demand pooled transit feeder service. The final area ranking is conducted based on the total KPI rank score: areas with lower totals, and thus higher final rankings, are considered to have the greatest potential for successful service implementation.

Ride-pooling potential increases at low demand levels (up to $1\%$), with the first threshold reached in most areas at demand fractions of $0.1\%$–$0.2\%$. The second and third thresholds are generally met at around $2\%$, except for Area 10, which fails both, and Area 4, which misses the third threshold within $\alpha \in [0.001, 0.5]$. For Area 9, thresholds are exceeded at lower demand levels, even when demand is extended to $3\%$. Area 9 with Hub 9 'Kraków Mydlniki (PKP)' consistently achieves a higher vehicle hours reduction and occupancy.

This observation is confirmed using ranking data (Table \ref{tab2}). When ranking the areas, we analyse the minimum demand fraction required to achieve three performance thresholds, with top ranks assigned to areas needing the smallest demand fraction. For the first threshold, Areas 3, 7, and 9 rank highest ($\alpha = 0.001$), while Area 10 ranks last ($\alpha = 0.02$). For the second threshold, Area 9 leads ($\alpha = 0.005$), and Area 10 fails to qualify within $\alpha \in [0.1\%, 5\%]$. For the third, Area 9 ranks first ($\alpha = 0.007$), and Area 10 again falls short. The total ranking identifies Area 9 with Hub 9 'Kraków Mydlniki (PKP)' as the most promising candidate with a total score of 3. Area 3 with Hub 1 'Czerwone Maki P+R' ranks second with 5 points. In contrast, Area 4 with Hub 1 and Area 10 with Hub 10 'Dunikowskiego' rank last.

\begin{figure}
	\centering
		\includegraphics[width=\linewidth]{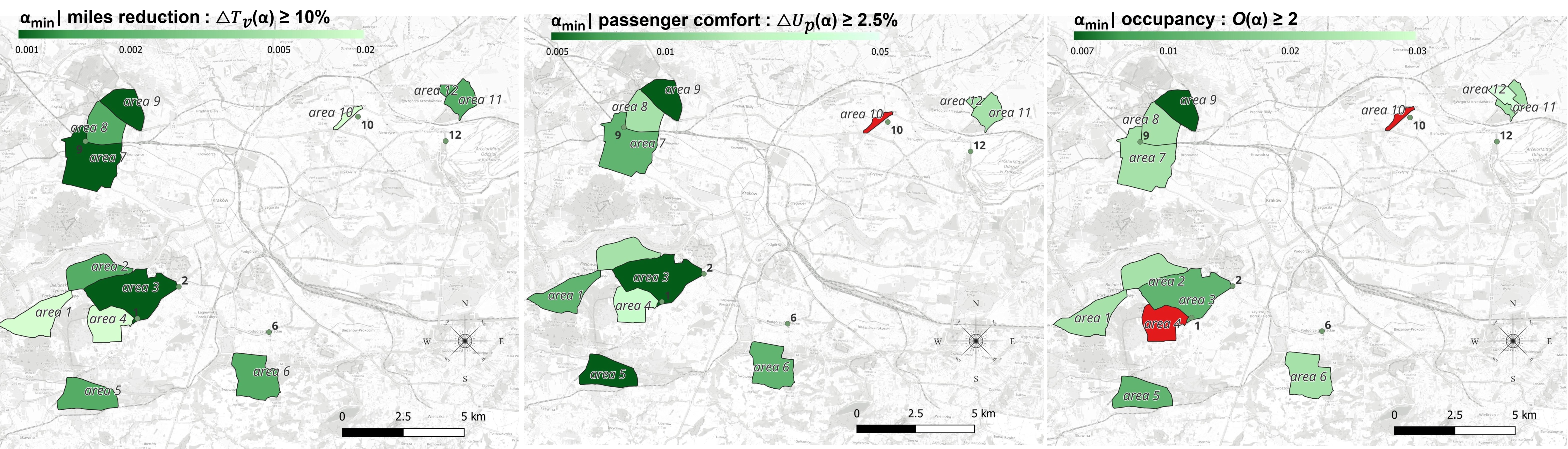}
	\caption{ Ranking of on-demand feeder service locations in Krakow: results for the first threshold are shown on the left, the second in the centre, and the third on the right. Areas that have not reached the thresholds within $\alpha\in[0.001,0.05]$ are indicated in red; the most promising areas are highlighted in dark green.}
	\label{FIG:8}    
\end{figure}

The Krakow map (Fig.~\ref{FIG:8}) shows the spatial distribution of areas and the minimum demand fractions needed to meet thresholds. Colour coding highlights promising areas (dark green) and less viable ones (red). Area 9 consistently ranks as the best candidate, marked in dark green on all thresholds. Area 10, marked red for the second and third thresholds, is the least suitable for service implementation.

\subsection{Benchmarks for the low-demand period}
\begin{figure}
	\centering
		\includegraphics[width=\linewidth]{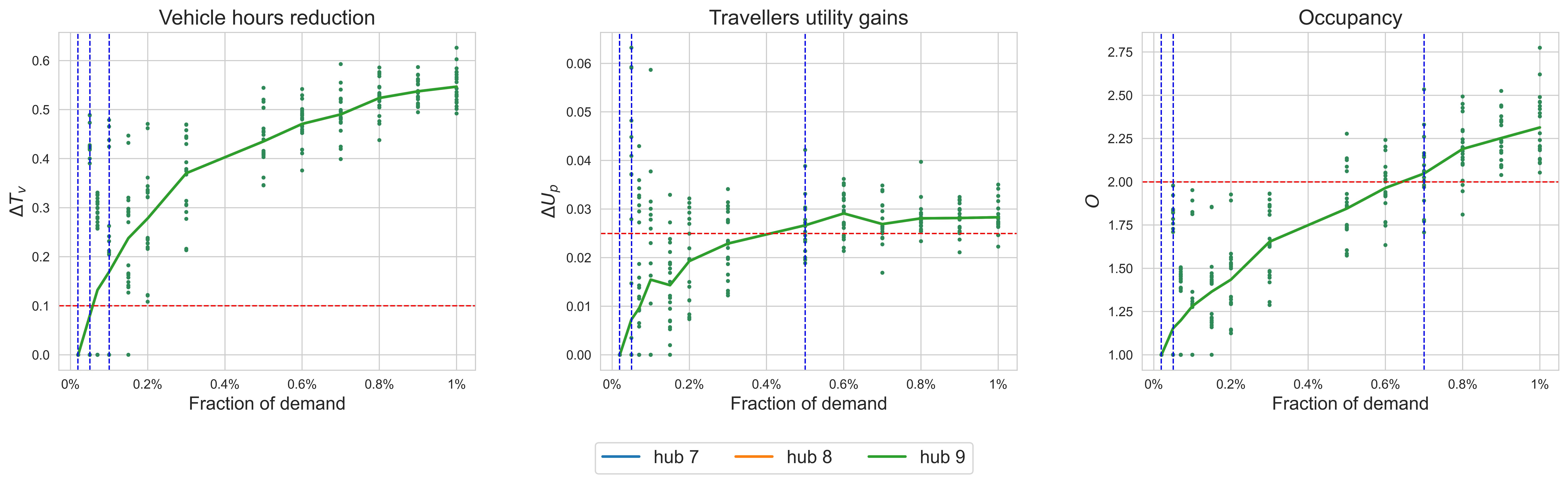}
	\caption{KPIs and three benchmarks for the combination of Area 9 and Hub 9 ’Kraków Mydlniki (PKP)’, plotted against demand levels. Horizontal dashed red lines indicate established KPI thresholds, while vertical blue lines mark three benchmarks.}
	\label{FIG:9}    
\end{figure}

For the promising Area 9, we examine three benchmarks across demand fractions, $\alpha\in[0,0.01]$, in 30 replications. Fig. \ref{FIG:9} presents KPIs for ride-pooling, illustrating early-stage service performance at various demand levels. Established thresholds are marked with dashed red lines, and the three key benchmarks are indicated by vertical blue lines. Each dot represents a single simulation result for a specific KPI at a given demand level, while the lines show average performance across replications.
The results show that pooled rides emerge at a demand fraction of $0.025\%$. Significant growth in ride-pooling potential occurs at $0.05\%$, with efficiency improving as more travellers use the service. The third benchmark is reached at demand fractions of $0.1\%$, $0.5\%$, and $0.7\%$ for the three KPIs, consistently meeting thresholds and indicating full operational capacity.
Identifying these benchmarks provides valuable information to municipalities and helps them understand when the service will become sustainable after launch.

\section{Discussion and conclusion}

We propose a parameter-free approach using demand fractions to simulate potential demand, allowing the evaluation of on-demand transit feeders in urban multimodal networks before implementation. This method helps municipal authorities identify the most promising areas for the launch of new services, even without exact demand data.
Our study focusses on selecting area-hub combinations to improve public transport accessibility in low-demand urban regions. The Krakow case study demonstrates the methodology’s effectiveness, identifying specific areas where on-demand feeder services are most viable, along with benchmarks for initial implementation. Three KPIs are assessed: vehicle kilometre reduction, passenger comfort, and occupancy at varying demand levels in 12 pre-selected areas, providing data-driven insights into the feasibility of new services.

The results identify Area 9, paired with Kraków Mydlniki (PKP), as the most promising candidate (Table \ref{tab2}). It should be noted that the minimum demand fractions identified in this study represent thresholds at which pooling becomes operationally viable, not predictions of actual uptake. For the service to be practically deployable, the real-world fraction of residents willing to use the on-demand feeder service must exceed these minimum thresholds. In contexts where on-demand mobility adoption is currently low, additional demand stimulation measure, such as integration with PT ticketing, subsidised fares, or targeted awareness campaigns, may be required before the pooling thresholds identified here become reachable in practice. Higher-ranking areas, such as Area 9 and Area 3, exhibit a favourable balance of population density, hub distance, and infrastructure suitability for pooled transit (Table \ref{tab1}). In contrast, lower-ranking areas, such as Area 10, with a high population density but close proximity to the Dunikowskiego hub, demonstrate the least potential, as the need for additional feeder services decreases with shorter distances to public transit options.

For the top-ranked Area 9, we establish three benchmarks to assess early-stage service performance (Fig. \ref{FIG:9}). Ride-pooling potential becomes significant at demand levels of $0.05\%$, with efficiency thresholds met at $0.1\%$, $0.5\%$, and $0.7\%$ for the three KPIs. These benchmarks guide municipal planning, enabling phased implementation aligned with demand growth projections.

Spatial analysis (Fig. \ref{FIG:8}) highlights variations in ride-pooling feasibility across Krakow. Area 9, marked in dark green for all thresholds, shows a high likelihood of meeting service benchmarks even at low demand levels. In contrast, areas like Area 10 and Area 4, marked in red, are less promising, failing to meet certain thresholds due to proximity to existing hubs or other factors requiring further study.

The findings demonstrate the utility of a parameter-free approach for evaluating on-demand transit feeders. By balancing population density with transit hub accessibility, this method provides actionable insights to maximise ride-pooling efficiency and inform municipal decision-making.

\subsection{Limitations and future works}

Despite its merits, our study has certain limitations. The ExMAS algorithm is limited to point-to-point ride-hailing, assessed only in comparison to solo ride-hailing. The ride-hailing baseline sets an upper bound on the demand available for pooling. In practice, the real-world fraction of residents willing to use the on-demand service may be smaller than the values of $\alpha$ tested here – some will walk, cycle, or drive to the hub regardless. Actual pooling gains may therefore be smaller than reported. Incorporating observed mode choice is a natural extension of the method once post-launch monitoring data are available. Additionally, demand must be predetermined, and the fleet is not explicitly managed. Introducing explicit fleet modelling, for instance following the dynamic trip-vehicle assignment framework of \citet{alonso2017demand}, would allow more realistic estimates of pooling rates under capacity constraints and remains a priority for future work.
The experiment was conducted in Krakow, a mid-sized European city, using a medium-scale sample. Our analysis focusses solely on the first-mile ride from pick-up points to hubs, simplifying the model, as the efficiency of on-demand pooled transit depends on the entire trip taken by travellers. Some areas, particularly those farther from the centre, may experience longer travel distances.
The analysis also assumes a single predetermined hub for all travellers leaving the area, without considering individual hub selection. Furthermore, the study only considers population potential, ignoring factors such as travel motivations, goals, current transportation habits, and existing public transport options. We assume that all travellers intending to use the system aim to reach the hub, and we focus on system operation parameters.
Future research should consider the entire journey, including both the feeder segment and the public transit segment, to provide a more comprehensive assessment of public transport attractiveness. Examining factors such as demographics and time-of-day variations could improve demand estimation and service predictability. Further studies could also test the scalability of this method in different urban contexts, allowing cross-regional comparisons and deeper insight into the role of on-demand feeder services in complementing traditional public transit.
By applying this methodology to various urban settings, we can identify universal patterns in shared mobility potential and inform the development of effective on-demand feeder services.

\textbf{Acknowledgements}

This research was supported by SUM project co-funded by the European Union’s Horizon Europe Innovation Action under grant agreement No. 101103646. Views and opinions expressed are however those of the author(s) only and do not necessarily reflect those of the European Union or of the European Climate, Infrastructure and Environment Executive Agency (CINEA). Neither the European Union nor the granting authority can be held responsible for them.

\textbf{Author contribution}

O.S.: Investigation, Writing - Original Draft. 
H.V.: Software.
M.B.: Software, Writing - Reviewing and Editing.
F.G.: Software.
R.K.: Supervision, Conceptualisation, Writing - Reviewing and Editing. 



\textbf{Data availability}

Code, sample data, and raw results are available in the public repository \url{https://github.com/OlhaShulikaUJ/SUM_project/tree/main/NSM}.
Due to GDPR constraints, address point data is not shared, but sample synthetic data is provided for illustration.

\section*{Declarations}

\textbf{Competing interests}

The authors declare no competing interests.



=============

\bibliography{sn-bibliography}

\end{document}